%% file: am2003-proc.tex
\documentclass[12pt]{article}
\usepackage{graphicx}    % standard LaTeX graphics tool
\include{definitions}

\begin{document}

\title{
Some aspects of radiative corrections and non-decoupling effects of heavy Higgs bosons in two Higgs Doublet Model
}
%titlerunning{Non-decoupling efects of heavy Higgs bosons in THDM}
\author{Michal Malinsk\'y
\thanks{
	Institute of Particle and Nuclear Physics, Charles University, Prague; S.I.S.S.A., Trieste
	 \newline 
	 \mbox{} \hskip 6mm E-mail: {\tt malinsky@ipnp.troja.mff.cuni.cz}
	}
}
\maketitle
\abstract{The possibility of having relatively large non-decoupling effects of the heavy Higgs particles within the two-Higgs doublet extension of the electroweak standard model is briefly discussed and demonstrated on an example of the one-loop amplitude of the process $e^+e^-\to W^+W^-$.}

\section{Introduction}
\label{Introduction}
Although the two-Higgs-doublet extension of the standard model (THDM) was invented about 30 years ago \cite{Lee:iz}, it still belongs among viable candidates for a theory beyond the electroweak standard model (SM). Despite its simplicity it is quite popular, namely because of its capability to include various aspects of ``new physics'' like for example the additional sources of CP violation (see e.g. \cite{Krawczyk:2002df}, \cite{Iltan:2001gf}). 
Moreover, its two Higgs doublet structure mimics many features of the Higgs sector of perhaps the most popular SM extension, the minimal supersymmetric standard model (MSSM). On the other hand, since the Higgs sector of THDM is less constrained, it can lead to various effects which are not present in MSSM, in particular to the non-decoupling behaviour of the heavy Higgs boson contributions in the electroweak scattering amplitudes. 

%\subsection{Basics of THDM}
%\label{THDMbasics}
As in the MSSM, the presence of the additional doublet leads to five physical Higgs states: 2 CP even Higgs scalars $h^0$ and $H^0$, a CP odd pseudoscalar $A^0$ and a charged pair of $H^\pm$. The  lightest scalar $h^0$ is quite similar to the SM Higgs boson $\eta$ i.e. the mass of $h^0$ should be close to the weak scale. On the other hand the typical mass scale of the other Higgses ($M_H$) is not so constrained in general, the unitarity bounds \cite{Akeroyd:2000wc} permit $M_{H}$ around one TeV (if there is no new physics in the game at this scale). Therefore a natural question arises as to whether these additional Higgs bosons  tend to decouple from the weak-scale amplitudes. As we shall see in the next section, the answer is 'not in general'. 

\section{Non-decoupling of heavy Higgs bosons in THDM}
The reason why the heavy Higgs bosons {\it need not} decouple from the weak-scale physics in the THDM, but they do so within MSSM \cite{Dobado:2000pw} is roughly the following. Since the Higgs self-couplings are driven by SUSY, the only way to make the four additional Higgs bosons ($H^0$, $A^0$ and $H^\pm$) sufficiently heavy in MSSM  is to adjust the $SU(2)_L\otimes U(1)_Y$ singlet mass parameters in the Higgs potential; in such case these masses have to decouple in accord with the famous Appelquist-Carazzone theorem \cite{Appelquist:tg}. In the THDM case one can do the job also by a convenient choice of the Higgs couplings $\lambda_i$ and the SSB parameter $\tan \beta$, keeping at the same time the singlet mass parameters small. Notice that even the violation of the simple unitarity bounds could be fully compatible with the requirement of perturbativity of the Higgs sector ($\lambda_i < 1$) provided one can choose a sufficiently large value of $\tan\beta$. 
As an illustration, consider the following tree-level THDM Higgs mass relations: 
\begin{eqnarray}\label{masses}
m_{h^0}^2 & = & \frac{1}{2}(1-\kappa)M^2+\left[B_2\sin^2\beta - A_1\cos^2\beta + \frac{1}{4}C (1+\cos 2\alpha \cos 2\beta)\right]\frac{v^2}{\cos 2\alpha }\nonumber \\
m_{H^0}^2 & = & \frac{1}{2}(1+\kappa)M^2+\left[A_2\cos^2\beta -B_1\sin^2\beta - \frac{1}{4}C (1-\cos 2\alpha \cos 2\beta)\right]\frac{v^2}{\cos 2\alpha } \nonumber \\
m_{A^0}^2 & = & M^2 -\frac{1}{2}\left[2\lambda_5^R+\lambda_6^R\cot\beta + \lambda_7^R\tan\beta\right] v^2  \\
m_{H^\pm}^2 & = & M^2 -\frac{1}{2}\left[\lambda_4+\lambda_5^R+\lambda_6^R\cot\beta + \lambda_7^R\tan\beta\right] v^2 \nonumber 
\end{eqnarray}
where (using the superscript $R$ to denote the real part of a quantity)
\begin{eqnarray}
M^2 \equiv  \frac{{m^2_{12}}^R}{\sin\beta \cos\beta} &\qquad\qquad & \kappa \equiv -\frac{\cos 2\beta}{\cos 2\alpha}\nonumber \\
A_1 \equiv  \lambda_1 \sin^2\alpha - \lambda_7^R \tan\beta \cos^2\alpha  & & B_1  \equiv  \lambda_2 \sin^2\alpha - \lambda_6^R \cot\beta \cos^2\alpha \nonumber \\
A_2 \equiv  \lambda_1 \cos^2\alpha - \lambda_7^R \tan\beta \sin^2\alpha  & & B_2  \equiv  \lambda_2 \cos^2\alpha - \lambda_6^R \cot\beta \sin^2\alpha\nonumber \\
C  \equiv \lambda_7^R \tan\beta - \lambda_6^R \cot\beta & & D  \equiv \lambda_7^R \tan\beta + \lambda_6^R \cot\beta \nonumber
\end{eqnarray}
The model and notation are those used in \cite{Malinsky:2003bd}. Notice that in the case $\lambda_6=\lambda_7=0$ one recovers the relations obtained in \cite{Kanemura:1999xf}. 
Moreover, using
\begin{equation}
\cos^2(\alpha-\beta)=\frac{m_{h^0}^2-m_L^2}{m_{H^0}^2-m_{h^0}^2} \quad {\rm and}\quad m_L^2\equiv \lambda_1\cos^4\beta+\lambda_2\sin^4\beta+\frac{1}{2}\left(\lambda+2D\sin^2\beta\right)v^2
\end{equation}
it is easy to se that if the weak-scale contributions (square-brackets in (\ref{masses})) are small compared to  $M^2$, the requirement of having $h^0$ light and the others much heavier forces the heavy multiplet to be almost degenerate with masses proportional to $M$, which is the signature of the so-called decoupling regime \cite{Gunion:2002zf}. Therefore, it is the {\it distortion of the heavy Higgs spectrum} which matters concerning the possible nondecoupling effects of the additional Higgses in THDM. 

In this work I would like to demonstrate these issues at the particular case of the amplitude of the proces $e^+e^-\to W^+W^-$ at one-loop level in THDM in comparison with the well-known one-loop SM result \cite{Bohm:1987ck}. 
Note that there are already earlier papers on this topic in the literature \cite{Kanemura:1997wx},\cite{Malinsky:2002mq} but these usually make use of some specific approximations (in particular, the equivalence theorem for longitudinal vector bosons \cite{Cornwall:1974km}) which we would like to avoid.

\section{The process $e^+e^-\to W^+W^-$}
For the considered process, the central quantity of our interest is the deviation of the differential cross-section, calculated within THDM, from its SM value; this is defined by 
\begin{equation}
\delta \equiv {{\rm d}\sigma^{THDM}}/{{\rm d}\sigma^{SM}}%[e^+e^-\to W^+W^-]
-1
\end{equation}       
Expanding the THDM amplitude around the SM value and keeping just the leading terms, one gets \cite{MalinskyHorejsi2} 
\begin{equation}
\delta \doteq 2 {\rm Re} \frac{\Delta {\cal M}[\Delta \Gamma^{TGV}]_{1-loop}}{{\cal M}^{SM}_{tree}}
\end{equation}   
Here $\Delta {\cal M}_{1-loop}$ stands for the difference of the THDM and SM 1-loop amplitudes, which descends primarily (the leading term) from the differences of the triple gauge vertex corrections $\Delta \Gamma^{TGV}$:
$$
\Delta {\cal M}[\Delta \Gamma^{TGV}]_{1-loop} = 
    \left[\parbox{2cm}{
		\includegraphics[height=1.3cm, width=2cm]{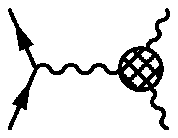}}
    \right]_{THDM} -
    \left[\parbox{2cm}{
		\includegraphics[height=1.3cm, width=2cm]{./eps/tgv_amp.eps}}
    \right]_{SM}
$$
%\begin{equation}
%= \sum_{{V}=\gamma,{Z}} \bar{v}(p_1)\gamma^\sigma 
%u(p_2)\frac{g_{ee{V}}g_{VWW}\Delta\Gamma^{V}_{\sigma\mu\nu}}
%{s-m^2_{V}}
%\,\varepsilon^{*\mu}(q_1)\varepsilon^{*\nu}(q_2)
%\end{equation}
%\subsection{One-loop triple gauge vertices in THDM}
Since most of the technical aspects of the calculation are covered in \cite{Malinsky:2003bd},\cite{MalinskyHorejsi2} let us emphasize only several salient points. {\it i)} We have chosen to work in the on-shell renormalization scheme. There are two main reasons for that: the overall number of diagrams to be calculated is reduced with respect to other schemes and the mass-parameters we are playing with are the true physical masses. The only disadvantage is the need of treating carefully the finite parts of the counterterms which must be computed by means of Ward identities. On the other hand, the cancellation of UV-divergences provides a non-trivial consistency check. {\em ii)} There is also a simple consistency check for the finite parts of $\Delta \Gamma^{TGV}$: they should tend to vanish in the decoupling regime, i.e. in the case where the masses of heavy Higgs bosons are large and almost degenerate.        

\section{Summary of results and conclusion}
\begin{figure}
\center
\includegraphics[height=5.6cm, width=10cm]{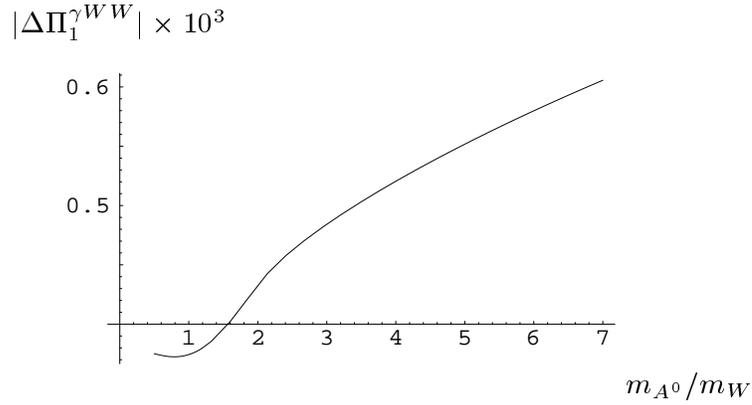}
%\picplace{5cm}{2cm} % Give the correct figure height and width in cm
%
\caption{$|\Delta \Pi_1^{\gamma WW}|$ as a function of $m_{A^0}$. The other masses are: $m_\eta = 105$GeV, $m_{h^0} = 125$GeV, $m_{H^0} = 145$GeV, $m_{H^\pm} = 180$GeV and we take $\sqrt{s}=250$GeV}
\label{fig:1}       % Give a unique label
\end{figure}

\begin{figure}
\center
\includegraphics[height=5.6cm, width=10cm]{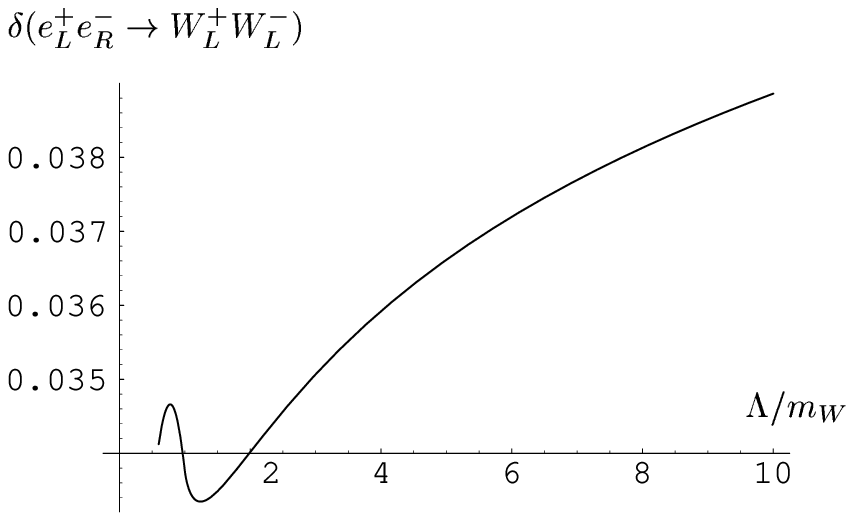}
%\picplace{5cm}{2cm} % Give the correct figure height and width in cm
%
\caption{$\delta$ as a function of $m_H^0=20\Lambda$, $m_{A^0}=10\Lambda$, $m_{H^\pm}=2\Lambda$.   $\sqrt{s}=320$GeV. However, for large $\Lambda$ the unitarity bounds can be violated.}
\label{fig:2}       % Give a unique label
\end{figure}

Due to the large number of diagrams contributing to $\Delta \Gamma^{TGV}$ it is hard to get an analytic expression even for the leading terms in $\Delta {\cal M}_{1-loop}$. The numerical analysis shows that the formfactors   $\Delta \Pi^{\gamma WW}$ and $\Delta \Pi^{ZWW}$ defined in \cite{Malinsky:2003bd} behave in accordance with the consistency conditions mentioned above. For example, let us look at  $|\Delta \Pi_1^{\gamma WW}|$ as a function of the mass of the $A^0$, (fig.\ref{fig:1}): since the other Higgs masses are kept close to the weak scale, the heavy Higgs spectrum distortion grows with $m_{A^0}$ and the non-decoupling effect in the formfactor as well. 
Concerning $\delta$, one naturally expects a similar behaviour because it is linear in the formfactors (at the leading order, see \cite{Malinsky:2003bd}). Let us take the particular case: $e^+_Le^-_R\to W^+_L W^-_L$ (in this setup the leading term turns out to be $\cos\theta^*$-independent which allows us to draw simpler pictures). 
As can be seen in fig.\ref{fig:2}, for large distortions of the heavy Higgs spectrum one can get an effect of several percent.
At least in principle, the nondecoupling effects of relatively heavy additional Higgs degrees of freedom can be used in an indirect exploration of the EW Higgs sector at future colliders. 
\\ 
\\
\noindent
{\bf Acknowledgements:}
The work was partially supported by "Centre for Particle Physics", project No. LN00A006 of the Czech Ministry of Education. 
I would like to thank Prof. Ji\v{r}\'\i\,  Ho\v{r}ej\v{s}\'\i\, for useful discussions. I am grateful to S.I.S.S.A. for the financial support and the organizers for the possibility to take part at this nice event.  

\end{document}

%% file: definitions.tex
\def\beq{\begin{equation}}
\def\eeq{\end{equation}}

\def\bea{\begin{eqnarray}}
\def\eea{\end{eqnarray}}
\def\ba{\begin{array}}                  %array
\def\ea{\end{array}}